\documentclass[twocolumn]{aastex631}

\usepackage{amsmath}
\usepackage{bm}
\usepackage{diagbox}\usepackage{tabularx}
\usepackage{makecell}
\usepackage{xcolor}
\usepackage{bibentry}

\graphicspath{{./}{Figures/}}

\begin{document}
\title{How Land-Mass Distribution Influences the Atmospheric Dynamics of Tidally Locked Terrestrial Exoplanets}

\author{F. Sainsbury-Martinez}
\affiliation{School of Physics and Astronomy, University of Leeds, Leeds LS2 9JT, UK}
\author{C. Walsh}
\affiliation{School of Physics and Astronomy, University of Leeds, Leeds LS2 9JT, UK}
\author{G. J. Cooke}
\affiliation{Institute of Astronomy, University of Cambridge, UK}
\affiliation{School of Physics and Astronomy, University of Leeds, Leeds LS2 9JT, UK}
\author{D. R. Marsh}
\affiliation{School of Physics and Astronomy, University of Leeds, Leeds LS2 9JT, UK}

\begin{abstract} 
Interpretation of the ongoing efforts to simulate the atmospheres of potentially-habitable terrestrial exoplanets requires that we understand the underlying dynamics and chemistry of such objects to a much greater degree than 1D or even simple 3D models enable. Here, for the tidally-locked habitable-zone planet TRAPPIST-1e, we explore one effect which can shape the dynamics and chemistry of terrestrial planets: the inclusion of an Earth-like land-ocean distribution with orography. To do this we use the Earth-system model WACCM6/CESM2 to run a pair of TRAPPIST-1e models with N$_2$-O$_2$ atmospheres and with the sub-stellar point fixed over either land or ocean.  The presence of orography shapes atmospheric transport, and in the case of Earth-like orography, breaks the symmetry between the northern and southern hemispheres which was previously found in slab ocean models. For example, peak zonal jet speeds in the southern hemisphere are $50\textrm{---}100\%$ faster than similar jets in the northern hemisphere. This also affects the meridional circulation, transporting equatorial material towards the south-pole. As a result we also find significant changes in the atmospheric chemistry, including the accumulation of potentially lethal quantities of ozone at both the south pole and the surface. 
Future studies which investigate the effects of land-mass distribution on the dynamics of exoplanetary atmospheres should pay close attention to both the day-side land-fraction as well as the orography of the land. Simply modelling a flat land-mass will not give a complete picture of its dynamical impact. 
\end{abstract}


\section{Introduction} \label{sec:introduction}

The search for a habitable terrestrial exoplanet remains a tricky prospect, with the transit signal associated with the atmosphere of an Earth-analogue planet passing in front of a Sun-like star lying below the current detection limits of both space and ground-based spectrometers. Although future space missions, such as LIFE \citep{2022A&A...668A..52K} or the Habitable Worlds Observatory \citep{2021pdaa.book.....N} may be able to characterise Earth-analogue atmospheres. However, planetary systems orbiting cooler stars, particularly M-dwarfs, offer a current-day solution to this conundrum. The transits of said planets are generally easier to detect than their G-dwarf orbiting counterparts thanks to the significantly smaller radius of M-dwarfs increasing the relative signal of both a transiting planet and its extended atmosphere. Furthermore, the lower luminosities of these stars mean that the habitable zone, which is broadly defined as the region around a star in which the equilibrium temperature of a planet would fall into the range that allows for liquid water on the surface, lies much closer to the host star \citep{2013ApJ...765..131K}. 
For example, the habitable zone of TRAPPIST-1 lies between $\sim$0.025 and $\sim$0.05 au \citep{refId0,2016Natur.533..221G}, which corresponds to an orbital period of between $\sim4.5$ and $\sim13.5$ days. This means that that multiple transits of any potentially habitable planets (such as TRAPPIST-1e, the most likely of the habitable zone TRAPPIST-1 planets to host a potentially terrestrial atmosphere - \citealt{2017ApJ...839L...1W,2021ApJ...913..107K}) can be combined, reducing the signal-to-noise ratio and allowing for the detection of atmospheric constituents within weeks/months rather than years/decades \citep{Lustig-Yaeger_2019}.  \\

However, the proximity of such planets to their host star is also likely to have implications for their orbital dynamics. Specifically, the small orbital distance between habitable-zone planets and their host M-dwarf star means that angular momentum exchange via tidal torques between the two bodies is likely to lead to the synchronisation of the planetary rotation rate and orbital period. This is known as tidal locking and it implies a permanently illuminated day-side and a permanently dark, and hence cooler, night-side \citep{1964hpfm.book.....D,2017CeMDA.129..509B}.
This is likely to have significant implications for the atmospheric dynamics, with the strong day-night temperature/pressure gradient acting as one of the primary drivers of the global circulation. 
The other driver of the atmospheric dynamics is, of course, the somewhat rapid planetary rotation ($P_{rot}<15$ days), with the Coriolis effect playing a significant role in shaping any off-equator circulations. On the Earth these effects include the geostrophic winds that drive the extra-tropical cyclones/anti-cyclones that are responsible for much of the US's and Europe's weather \citep{OntheExistenceofStormTracks,StormTrackDynamics,wcd-2-1111-2021}. \\
Taken together, the fixed stellar insolation and the strong off-equator Coriolis effect can lead to the formation of standing Rossby and Kelvin waves which pump eastwards angular momentum from high-latitudes towards the equator, a process that can result in the formation of (a) super-rotating jet(s) \citep{2011ApJ...738...71S}. Such a process has been predicted \citep{2002A&A...385..166S} and observed \citep{2007Natur.447..183K,Zellem_2014} for hot Jupiters (Jupiter-sized planets which orbit close to their host star) and is also expected to play an import role in the atmospheric dynamics, and chemistry, of terrestrial exoplanets \citep[e.g.][]{2015MNRAS.453.2412C,2014MNRAS.445..930C,2015MNRAS.453.2412C}. \\

Recently, \citet{10.1093/mnras/stad2704} used an Earth-like Global Circulation Model (GCM) of Proxima Centauri b to suggest that a stratospheric dayside-to-nightside overturning circulation should advect ozone from its formation location on the day-side to the night-side. Here ozone persists due to a combination of a lack of UV irradiation to drive photolytic loss processes and confinement in off-equator gyres associated with global standing Rossby and Kelvin waves. However the hemispherically-symmetric winds found by \citet{10.1093/mnras/stad2704} are likely only possible thanks to their use of a global slab ocean (i.e. a motionless ocean which is assumed to be well mixed and covers the entire planetary surface) instead of a dynamic ocean or accounting for the presence of land-masses. { The inclusion of a dynamic ocean can significantly alter heat transport, including day-night heat transport, thus affecting the day-night temperature contrast and hence the strength of the global overturning circulation \citep{2014PNAS..111..629H}. The inclusion of landmasses can break global wind symmetries (as on Earth) and affect the circulations found in a dynamic ocean \citep{2020ApJ...896L..16S}. A good example of the effect that land-masses can have on the global symmetries of atmospheric circulations is \citet{2019AsBio..19...99D}, who explored models of Proxima Centauri b with Earth-like land-mass distributions and found that the topography reshaped winds, breaking symmetries including the location of (Rossby) gyres. However they did not include a coupled chemistry scheme in their model, and hence were unable to investigate, for example, the effects of landmasses on the ozone distribution.  } 
{ Since then, such a model has been run:} \citet{anand2024} showed that, when including an Earth-like land-ocean distribution in a model of the tidally-locked terrestrial exoplanet TRAPPIST-1e, any symmetry in the ozone distribution goes away. Instead the ozone accumulates over the south pole due to a combination of the same overturning circulations as \citet{10.1093/mnras/stad2704} transporting ozone from the day-side to the night-side in the outer atmosphere, and an asymmetry in the near-surface winds on the night-side transporting material southward. They attribute the asymmetry in the near-surface winds to wave-breaking associated with the land-ocean boundaries (for more details of these effects see, for example, \citealt{1992JCli....5.1181B,2019npCAS...2...10S,2022RvGeo..6000730P,https://doi.org/10.1029/2020RG000730}). Note that a similar result was found by \citet{cooke2024}, who discuss how the cold night-side allows for the accumulation of potentially lethal ($>40$ppbv) levels of surface ozone on tidally-locked exoplanets.  \\

In this work we explore how the presence of Earth-like orography affects the atmospheric dynamics of terrestrial exoplanets in more detail, including investigating if the global atmospheric circulations (and resulting advection of ozone) are affected by the land-ocean distribution at the sub-stellar point. \\
In \autoref{sec:method} we introduce our model, WACCM6/CESM2, an Earth System Model which has been used to study the atmospheres of both Earth-analogue \citep{10.1093/mnras/stac2604,2023MNRAS.524.1491L} and tidally-locked \citep{2023ApJ...959...45C,anand2024,cooke2024} exoplanets with pre-industrial Earth-like atmospheric compositions and land-mass distributions. Here we use WACCM6/CESM2 to run two models of TRAPPIST-1e, one in which the sub-stellar point is fixed over the pacific ocean and one in which it is fixed over land (specifically central Africa). { Note that this pair of relative land-mass distributions is the same as was considered by \citet{2019AsBio..19...99D}, allowing us to, much like them, explore if the presence of a land-mass at the substellar point has a significant effect on the atmospheric dynamics and, using our coupled model, chemistry. }In \autoref{sec:results} we compare and contrast the aforementioned models in more detail. This includes exploring differences in global atmospheric chemistry, differences (and similarities) in both zonal and meridional flows, and how orography shapes the horizontal winds by acting as a local source of divergence and vorticity. We also investigate how the winds shape the ozone distribution, driving the accumulation over the south pole (Antarctic) found by both \citet{anand2024} and \citet{cooke2024}. We finish, in \autoref{sec:concluding_remarks}, with some concluding remarks, discussing the implications of our results as well as the need for future model development of more flexible land-mass and orography models which can be coupled with models like WACCM6/CESM2 in order to better understand how such effects, and the associated symmetry breaking, might shape future observations of the atmospheres of potentially habitable exoplanets.

\section{Method: Modelling TRAPPIST-1e with WACCM6/CESM2} \label{sec:method}

To understand how the land-ocean distribution can shape the winds and resulting atmospheric chemistry of tidally-locked terrestrial exoplanets, we explore the atmospheric dynamics found in two tidally-locked TRAPPIST-1e models based upon the work of \citet{2023ApJ...959...45C}. These models are based on a version of WACCM6/CESM2 which has been modified to account for synchronous rotation\footnote{\url{github.com/exo-cesm/}} (i.e. a tidally locked planet in which the location of the incoming stellar insolation is fixed). \\

WACCM6 is a well-documented \citep{https://doi.org/10.1029/2019JD030943} high-top (the atmosphere extends to $\sim140$ km above the surface) configuration of version 2 of the Coupled Earth System Model (CESM2). It includes a modern, Earth-like, ocean and land model (including orography), and an initial atmospheric composition which approximates the pre-industrial Earth. { That is to say an Earth-like atmosphere, primarily composed of nitrogen and oxygen, with smaller amounts of water vapour, methane, carbon-dioxide and lighter noble gases. It does not include any human induced changes, such as polution or greehouse gas enhancement. For instance, the atmosphere is 285 ppm CO$_2$ vs the present day value of $>420$ ppm.}
Horizontally, both simulations have a resolution of 1.875$^\circ$ by 2.5$^\circ$ (corresponding to 96 cells latitudinally and 144 cells longitudinally), whilst vertically the simulation domain is split into 70 atmospheric levels distributed between 1 and $4.5\times10^{-9}$ bar, in $\log(P)$ space, such that the number of pressure levels increases near the surface (where the atmosphere is more dynamically active). Finally both models have been integrated for over 300 years (with a 30 minute timestep) in order to ensure that any effects associated with the atmospheric dynamics settling into a state associated with synchronous rotation have dissipated. { For example, we find no long term trends in the atmospheric or surface temperature.}  For our analysis, we consider a temporal average over the last 30 years of simulation time. A more in-depth discussion of WACCM6/CESM2, including the chemistry, radiation, and cloud physics, can be found in \citet{2023ApJ...959...45C} and \citet{2023MNRAS.524.1491L}. { More details on the Earth-tuned chemical network can be found in \citet{2019JGRD..12412380G} and \citet{2019MS001882}.}  \\

\begin{figure}[htp]
\begin{centering}
\includegraphics[width=0.99\columnwidth]{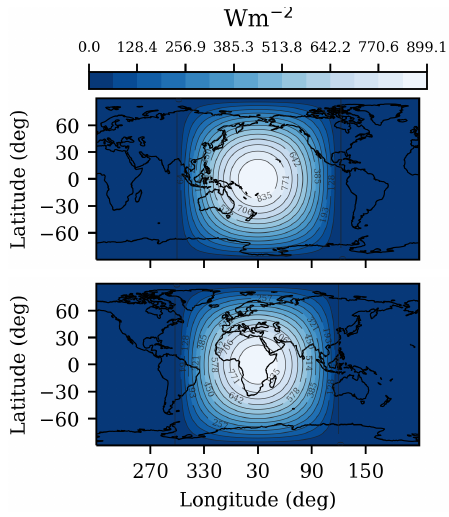}
\caption{ Maps showing the stellar insolation profiles for the two models considered here: one with the sub-stellar point placed over the pacific ocean (SSPO - top), and one with the sub-stellar point placed over land, specifically central Africa (SSPL - bottom). In both maps the land-mass distribution is outlined in black. \label{fig:stellar_insolation_maps}}
\end{centering}
\end{figure}

TRAPPIST-1e is a terrestrial planet which remains a strong object of interest in the ongoing search for a habitable, terrestrial, exoplanet. It has a radius of 0.91 R$_\earth$ and an orbital period of only $\sim6.1$ days. However, because TRAPPIST-1 is a cool M-dwarf, the peak insolation is close to that received by the Earth at 900 W m$^{-2}$ (around 66\% of that received by the Earth, but 50\% more than Mars), placing the planet at the cooler-edge of the habitable zone (see \autoref{tab:TP1E_parameters} for more details).  To match this peak insolation, we rescale the TRAPPIST-1 spectrum of \citet{2019ApJ...871..235P} (calculated using the PHOENIX stellar atmospheric code \citep{HAUSCHILDT1993301,2006A&A...451..273H,2007A&A...468..255B}) following the methodology of \citet{2023ApJ...959...45C}. { That is to say we rescale the integrated flux to match that received by TRAPPIST-1e whilst also rebinning the spectrum onto the grid used by WACCM6/CESM2}. \\
The resulting insolation maps can be seen in \autoref{fig:stellar_insolation_maps}, which also reveals the main difference between the two models considered here: one has the sub-stellar point fixed over the pacific ocean (henceforth referred to as SSPO), whilst the other has the sub-stellar point fixed over central Africa (henceforth referred to as SSPL). We select these two scenarios to examine which has a greater controlling effect on the global atmospheric dynamics of a tidally locked planet: the presence of land/ocean at the sub-stellar point, or differences in the land-mass distribution (and associated orography) between the northern and southern hemisphere. The latter scenario is of particular interest for an Earth-like land-mass distribution in which 68\% of the land can be found in the northern hemisphere, opening up the possibility of significant symmetry breaking between near surface flows in the northern and southern hemispheres. As previously discussed, and explored here, such asymmetries may play a key role in understanding differences between the latitudinally symmetric and asymmetric ozone distributions of \citet{10.1093/mnras/stad2704} and \citet{cooke2024,anand2024}. 

\begin{table}[htp]
\centering
\begin{tabular}{lrl}
Parameter & Value & Unit \\ \hline
Radius $R$ & 0.91 & R$_\earth$\\
Mass $M$  & 0.772 & M$_\earth$ \\
Semimajor Axis $a$ & 0.0292 & au \\
Orbital Period $P_{orb}$ & 6.099 & days\\
Obliquity $\epsilon$ & 0 & \, \\
Eccentricity $e$ & 0 & \, \\
Peak Insolation $I$ & 900 & W\,m$^{-2}$ \\
Surface Gravity $g$ & 9.1454 & m\,s$^{-2}$
\end{tabular}
\caption{Planetary parameters of TRAPPIST-1e, taken from \citet{10.1093/mnras/sty051}, \citet{2018A&A...613A..68G}, and \citet{2021PSJ.....2....1A}, with the final mass and radius of the planet being chosen such as to be consistent with the TRAPPIST-1 Habitable Atmosphere Intercomparison (THAI) program \citep{2022PSJ.....3..211T,2022PSJ.....3..212S,2022PSJ.....3..213F}. }
\label{tab:TP1E_parameters}
\end{table}

\subsection{The Helmholtz Wind Decomposition} \label{sec:helmholtz_method}
In order to investigate the horizontal wind dynamics in more detail, including how the surface drives wind asymmetries and fixed day-night forcing drives day-to-night transport, we turn to the Helmholtz wind decomposition, which has long been a staple of Earth atmospheric studies \citep{dutton1986ceaseless} and which has been applied to both hot Jupiters \citep{2021PNAS..11822705H} and tidally-locked terrestrial exoplanets \citep{2022PSJ.....3..212S}. \\
The Helmholtz wind decomposition splits the horizontal wind, $\bm{u}=\left(u,v\right)$ into two components: a divergent component which is `vorticity free' ($\bm{u}_{d}$) and a rotational component which is `divergence free' ($\bm{u}_{r}$). i.e.:
\begin{align}
  \bm{u} &= \bm{u}_{d} + \bm{u}_{r} \\
  &= -\bm{\nabla}\chi + \bm{k}\times\bm{\nabla}\psi,
\end{align}
$\bm{k}$ denotes a unit vector in the eastward/zonal direction, $\chi$ is the velocity potential function, $\psi$ is the velocity streamfunction, and both $\chi$ and $\psi$ can be linked to the divergence $\delta$ or vorticity $w$ directly:
\begin{align}
  \nabla^{2}\chi &= \delta\\
  \nabla^{2}\psi &= w.
\end{align}
The rotational component of the wind can be further split into eddy ($\bm{u}_{e}$) and zonal-mean ($\bm{u}_{z}$) components in order to isolate any zonal jets from rotational wind dynamics that they might be masking:
\begin{align}
  \bm{u}_{z} &= \left<\bm{u}_{r}\right>\\
  \bm{u}_{e} &= \bm{u}_{r} - \bm{u}_{z},
\end{align}
where $\left<\right>$ indicates the zonal-mean.\\
Hence each component of the Helmholtz wind decomposition probes a different part of the wind. For example, for a tidally-locked planet, $\bm{u}_{d}$ probes the global overturning circulation whilst $\bm{u}_{r}$ probes dynamics driven by angular momentum transport including zonal jets ($\bm{u}_{z}$) and the standing Rossby and Kelvin waves which drive them ($\bm{u}_{e}$). \\

\section{Results} \label{sec:results}

In order to investigate how differences in the land-mass distribution at the sub-stellar point, and between the northern and southern hemispheres, affects the atmospheric chemistry and dynamics of tidally-locked terrestrial (Earth-analogue) exoplanets, we start by examining the global atmospheric chemistry (\autoref{sec:chemistry}). We then focus most of our analysis efforts on the atmospheric circulations starting with the zonal-mean dynamics (\autoref{sec:zonal_meridional}) before moving onto horizontal wind (\autoref{sec:helmholtz}) and how the dynamics are shaped by orography (\autoref{sec:wave_breaking}). We finish with a discussion of the ozone distribution, focusing on the accumulation of ozone over the south pole due to orographically induced vortices (\autoref{sec:surface_ozone}).  

\subsection{Global Atmospheric Chemistry} \label{sec:chemistry}
\begin{figure*}[tp] %
\begin{centering}
\includegraphics[width=0.8\textwidth]{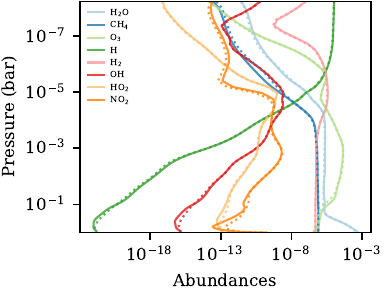}
\caption{ Comparison of the global and temporal mean fractional abundance profiles of eight potentially observable (i.e. H$_2$O, CH$_4$, O$_3$) or chemically important (i.e. H, H$_2$, OH, HO$_2$, NO$_2$) atoms and molecules. Profiles extracted from the model with the sub-stellar point placed over the ocean (SSPO) are shown as solid lines, whilst profiles extracted from the model with the sub-stellar point placed over land/Africa (SSPL) are shown as dashed lines. 
    \label{fig:abundance_comp} }
\end{centering}
\end{figure*}

\autoref{fig:abundance_comp} shows the global and temporal mean fractional abundance profiles of eight atoms/molecules which are either potentially observable (i.e. H$_2$O, CH$_4$, O$_3$) or a play a significant role in shaping the atmospheric chemistry (i.e. H, H$_2$, OH, HO$_2$, NO$_2$).  \\
Generally, the differences between the SSPO and SSPL models are small, and hence likely difficult to distinguish observationally. They can also be linked to a single primary driver: the relative influence of a liquid ocean at the sub-stellar point. As discussed in \citet{10.1093/mnras/stac1040,10.1093/mnras/stae554}, due to evaporation the fraction of the area around the sub-stellar point which is covered by oceans or land can have a significant effect on the water vapour content of the atmosphere. This is especially true for planets like TRAPPIST-1e (or as they consider, Proxima Centauri b) which not only receive less insolation from their host stars than the Earth does from the Sun, and hence are expected to be cooler, but are also tidally locked, which results in a cold, unirradiated, night-side. Together these factors can lead to a large fraction of their ocean away from the `hot' day-side being frozen \citep{2011ApJ...726L...8P,2019AsBio..19...99D,2022PSJ.....3..211T,2023ApJ...959...45C}. 
Indeed in both of our models we find that, away from the sub-stellar point, the ocean is entirely frozen over (not-shown), leaving our SSPO model in the so-called `Eyeball-Earth' state with a liquid ocean at the sub-stellar point, and our SSPL model with only narrow regions of liquid water near the coastline of Africa. \\
This difference in liquid ocean coverage leads to differences in the rate of evaporation of water from the surface, affecting/enhancing the overall water content of the atmosphere. Since this water evaporates on the day-side, it can then undergo photolysis, leading to an enhancement in both the oxygen and hydrogen content of the atmosphere. This oxygen can then go on to form other molecules, including (but not limited to) ozone, which may explain the enhanced ozone content that is hinted at in \autoref{fig:abundance_comp} (see the difference in ozone concentration both at low pressures, where it forms, and near the surface, around 0.1 bar), and seen in both the SSPO/SSPL models of \citet{cooke2024} and ourselves (\autoref{sec:surface_ozone}). \\
Overall it is the presence of this liquid ocean which drives many of the differences in atmospheric chemistry seen in \autoref{fig:abundance_comp}: the near surface differences are generally caused by the evaporation of water from the liquid ocean in the SSPO case (thus increasing the overall amount of hydrogen and oxygen in the atmosphere), whilst the high altitude changes, in particular the enrichment of oxygen-rich molecules, is due to water undergoing significant UV irradiation and photodissociating. But how does this water get here, and why do we find an enrichment in ozone near the surface for our SSPO model? 

\subsection{Zonal and Meridional Flows} \label{sec:zonal_meridional}
\begin{figure*}[htp]
\begin{centering}
\includegraphics[width=0.99\textwidth]{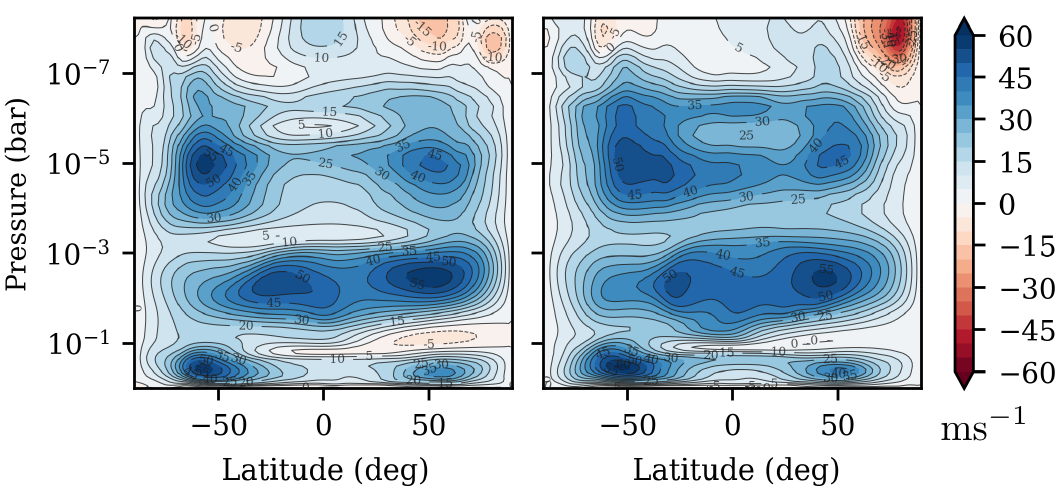}
\caption{Zonally and temporally averaged zonal wind profiles for both our model with the sub-stellar point placed over the ocean (SSPO - left) and sub-stellar point placed over land/Africa (SSPL - right). Here eastwards winds are shown in shades of blue whilst westwards winds are shown in shades of red. \label{fig:zonal_mean_zonal_wind}}
\end{centering}
\end{figure*}
\begin{figure*}[htp]
\begin{centering}
\includegraphics[width=0.785\textwidth]{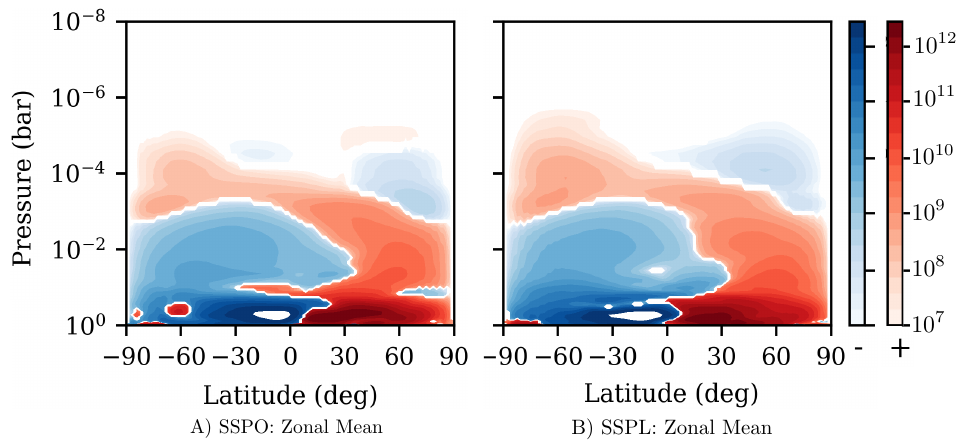}
\includegraphics[width=0.785\textwidth]{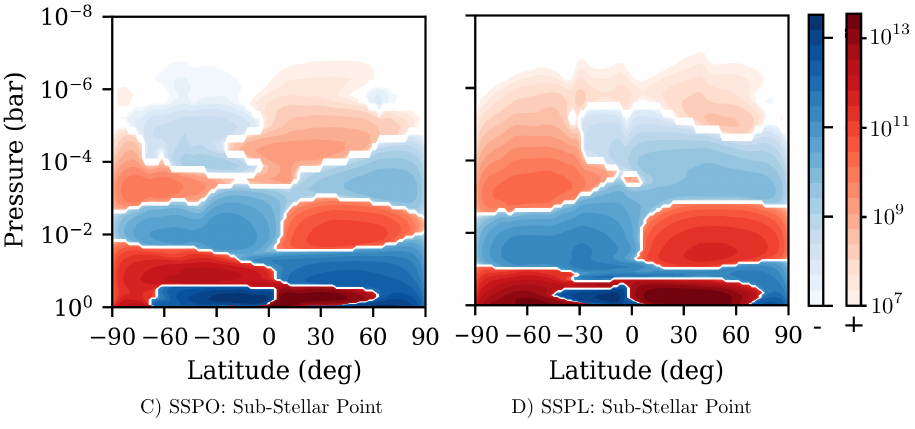}
\includegraphics[width=0.785\textwidth]{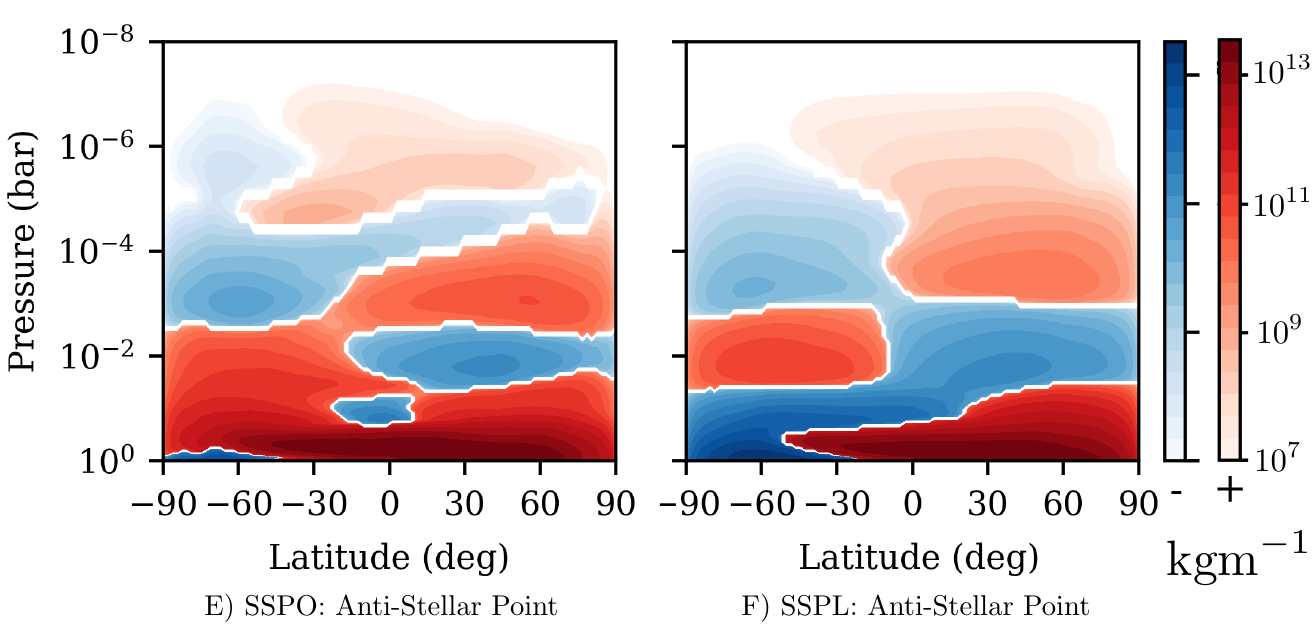}
\caption{ Select temporally averaged meridional circulation streamfunctions for both our model with the sub-stellar point placed over the ocean (SSPO - left) and sub-stellar point placed over land/Africa (SSPL - right). To demonstrate how the combination of tidally-locked thermal forcing and Earth-like orography shapes the atmospheric circulations, we show three different views of this circulation: the zonal-mean circulation (top), a $5^\circ$ average over sub-stellar longitudes (middle), and a $5^\circ$ average over anti-stellar longitudes (bottom). Note that the meridional circulation profile is plotted on a log scale with clockwise circulations shown in red and anti-clockwise circulations shown in blue. Thus, for example, in the zonal-mean circulation profiles, we find that the clockwise cell in the northern hemisphere and the anti-clockwise cell in the southern hemisphere combine to drive a near equatorial upflow at all pressures greater than $10^{-3}$ bar.   \label{fig:meridional_circulation}}
\end{centering}
\end{figure*}
To understand this transport, as well as the connection between the hot day-side and cold night-side, we next turn to the zonal wind (\autoref{fig:zonal_mean_zonal_wind}) and meridional circulation (\autoref{fig:meridional_circulation}). \\

\autoref{fig:zonal_mean_zonal_wind} shows the zonal-mean zonal-wind ($\left<u\right>$) for both our SSPO (left) and SSPL (right) cases. Here we can clearly see the similarities between the zonal winds in both models. We find three pairs of eastward zonal jets. In the outer atmosphere we find a pair of relatively symmetric jets, centred at $\sim10^{-5}$ bar ($\sim70$ km altitude) and $\sim\pm50^{\circ}$ latitude. Moving deeper, into the middle atmosphere, at a pressure of $\sim3\times10^{-3}$ bar ($\sim37$ km altitude) we find a pair of jets with a strong asymmetry in location between the southern ($-25^{\circ}$) and northern ($\sim50^{\circ}$) hemispheres. Finally, nearer the surface, at a pressure of $\sim 0.5$ bar ($\sim8$ km altitude) and a latitude of $\sim\pm50^{\circ}$ we find a pair of jets with a strong asymmetry in peak jet speed, with the southern hemisphere jet being almost twice as fast ($54\textrm{---}60$ m s$^{-1}$) as the jet found in the northern hemisphere ($31\textrm{---}34$ m s$^{-1}$).
The strong similarities between the zonal winds found in the SSPO and SSPL models, coupled with both models exhibiting a significant asymmetry between the northern and southern hemispheres, suggests that the primary driver of circulation asymmetries in our models is not the presence, or lack-thereof, of a land-mass as the sub-stellar point, but rather differences between dynamics in the northern and southern hemispheres. As we will discuss in \autoref{sec:helmholtz} and \autoref{sec:wave_breaking}, the difference between the two hemispheres that is most likely to be responsible is the land mass distribution (with 68\% of the land lying in the northern hemisphere) and the associated orography.  \\

Next, as shown in \autoref{fig:meridional_circulation}, we look at the meridional mass streamfunction (meridional circulation), investigating differences not only between our SSPO and SSPL model but also between the zonal-mean circulation and the localised circulation at the sub-stellar and anti-stellar points. \\
The meridional mass streamfunction takes the form:
\begin{equation}
    \psi=\frac{2\pi R_{p}}{g\cos\theta}\int^{P_{0}}_{P_{top}}v\,dP, \label{eq:meridional}
\end{equation}
where $v$ is the meridional velocity, $R_{p}$ is the radius of the planet, $g$ is the surface gravity, $\theta$ is the latitude, and $P_{0}$ and 
$P_{top}$ are the pressure at the surface and top of the atmosphere respectively. 
It describes the transport of material in the meridional plane (essentially a slice of the atmosphere taken at a single longitude, or averaged zonally when appropriate), and the interpretation of the figures is rather different from a wind map. Rather than the streamfunction representing flows directly it instead represents circulations. For example, in \autoref{fig:meridional_circulation}, clockwise circulations are shown in red, whilst anti-clockwise circulations are shown in blue. Where these circulations meet then represents net flows, either latitudinally or vertically. An example of this can be seen for pressures $>10^{-3}$ bar in the zonal-mean circulations (\autoref{fig:meridional_circulation}A/B). Here, the clockwise cell in the northern hemisphere and the anti-clockwise cell in the southern hemisphere combine to drive an upflow slightly north of the equator. \\
In general, we find that the aforementioned zonal-mean meridional circulation profiles are remarkably similar between our SSPO and SSPL models, with the main difference being the addition, in our SSPO model, of small clockwise circulations at the equator (at $P=10^{-1}$ bar) and south pole (at $P\simeq0.5$ bar), circulations which are not enough to overly affect the overall sense of the meridional transport, particularly at higher pressures. For pressures $>10^{-3}$ bar we find that material is transported downwards at the poles, equatorward at the surface, and generally upwards at the equator, although the symmetry is slightly broken and the upward transport tends to occur slightly off-equator in the northern hemisphere. Both aspects of this circulation are somewhat reminiscent of the Earth: the circulation cells resemble Hadley-cells (see, for example, the review of \citealt{atmos12121699}), albeit rather than extending from the equator to the tropics they extend all the way from the equator to the poles. The northward shift of the upflow away from the equator is similar to the offset seen in the inter-tropical convergence zone (i.e. Hadley-cell convergence zone - for more details see, for example, \citealt{WALISER2015121,itcz2022}). A similar single-cell-per-hemisphere structure was found by \citet{10.1093/mnras/stad2704}, although their circulation was symmetric about the equator which further reinforces our conclusion that the inclusion of an Earth-like land-mass distribution has significantly altered the dynamics between the northern and southern hemispheres. Such a conclusion would also explain why, as we move to lower pressures, the differences between our circulation and that of \citet{10.1093/mnras/stad2704} reduce, with both sets of models revealing a series of alternating clockwise and anti-clockwise circulations, likely associated with the strong thermal forcing of the outer atmosphere. { Finally, it has been found that rotation can play a role in shaping the extent of the Hadley cell. For example, \citet{1987JAtS...44..973D} found that, as you slow the rotation of an Earth-like planet, the latitudinal extent of the Hadley cell grows, approaching the poles for rotation periods $>10$ days. Similar results are also reported by, for example, \citet{Williams1988,1988ClDy....3...45W}, \citet{Navarra2002}, \citet{2014MNRAS.445..930C,2015MNRAS.453.2412C}, \citet{2018ApJ...852...67H} and \citet{2018GeoRL..4513213G} amongst others. Note however that, that the circulations found are generally symmetric about the equator, driven by a lack of symmetry breaking land-masses. } \\
Moving onto the circulation at the sub-stellar point (\autoref{fig:meridional_circulation}C/D), we find that, whilst differences between the SSPO and SSPL models have grown very slightly, the profiles still remain highly similar.
Near the surface we find that the circulation consists of two cells in each hemisphere: Hadley-like cells at low latitudes which drive a net-upflow at the equator, braced by Ferrel-like cells that extend from mid-latitudes to the poles. In the SSPO model the cells are somewhat symmetric between the northern and southern hemispheres, an effect which can be linked to the fact that we find ocean at almost all sub-stellar latitudes in this model. On the other hand, the role that the land-masses play in shaping the wind are much more apparent in the SSPL model, with the Hadley-Fellel cell transition occurring around the same latitude that land gives way to ocean in both the southern ($\sim30^{\circ}\textrm{---}\sim40^{\circ}$) and northern ($\sim70^{\circ}$) hemispheres (see \autoref{fig:stellar_insolation_maps}). 
Moving to lower pressure we find a stack of alternating clockwise and anti-clockwise cells, albeit with differences in the vertical extent of the cells between $10^{-3}$ and $10^{-1}$ bar, again likely due to differences in the wind induced by the land distribution and orography. Finally, at low pressures ($P\lesssim10^{-3}$) the circulation becomes highly time and longitude dependent, much more so than the low-pressure circulations on the night-side (see below), which suggests that the strong thermal forcing drives a highly dynamic atmosphere.  \\
A parallel story of similar circulations in the SSPO and SSPL models holds true near the anti-stellar point (\autoref{fig:meridional_circulation}E/F). Here we find a highly asymmetric circulation structure near the surface, with a single clockwise circulation cell extending from the north pole to a latitude of $\sim-50^{\circ}$ in the SSPL model and all the way to the south pole in the SSPO model (although the circulation also departs from the surface at a latitude of $\sim-50^{\circ}$). Overall, via a series of stacked circulation cells which extend to $\sim10^{-5}$ bar, we find that material will be transported from the outer atmosphere down towards the surface, where it will then be transported south towards the Antarctic. Furthermore, the differences between the near-surface transport found in the southern hemisphere of the SSPO and SSPL models may also explain the difference in peak ozone concentration found by \citet{cooke2024}. In the SSPL model, the southward surface transport does not extend all the way to the south pole, hence we must rely on a smaller anti-clockwise circulation cell to complete the poleward transport. This circulation may transport material aloft, reducing the overall concentration of ozone at the pole/surface.\\

\subsection{Helmholtz Wind Decomposition} \label{sec:helmholtz}
\begin{figure*}[tp]
\begin{centering}
\includegraphics[width=0.95\textwidth]{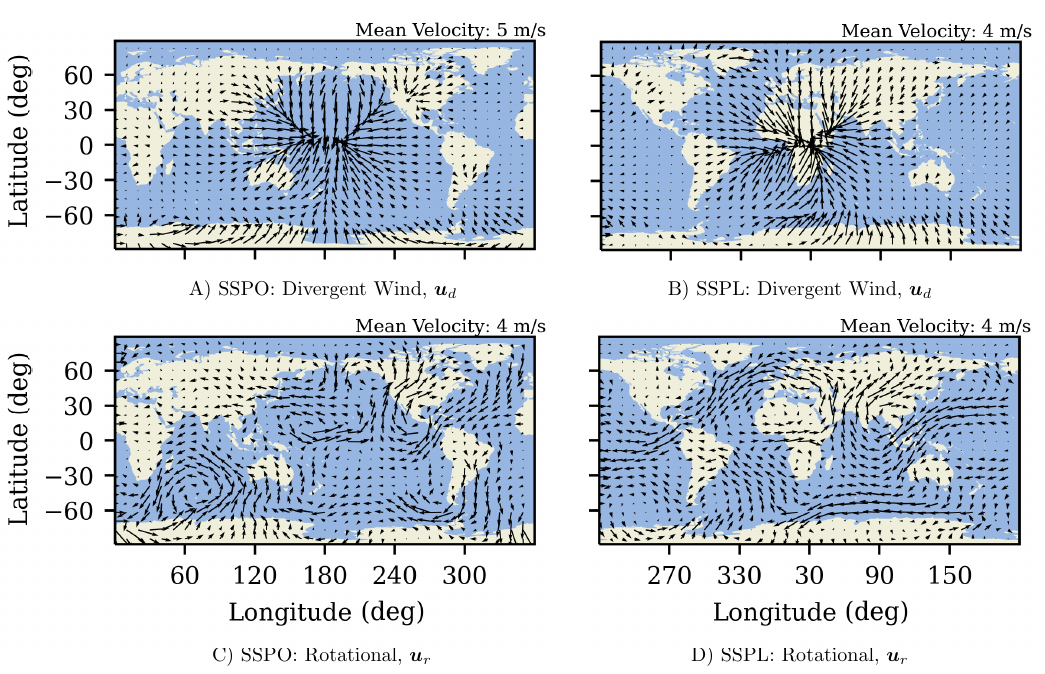}
\caption{Helmholtz decomposition of the radially averaged near-surface (over all $P>10^{-1}$ bar) horizontal winds for both our model with the sub-stellar point placed over the ocean (SSPO - right) and the sub-stellar point placed over land/Africa (SSPL - right). The top row plots the divergent component of the wind ($U_{d}$) whilst the bottom row plots the rotational component of the wind ($U_{r}$). Land-masses and oceans are shown in green/blue respectively. \label{fig:surface_Helmholtz_map}}
\end{centering}
\end{figure*}
\begin{figure*}[hpt]
\includegraphics[width=0.9\textwidth]{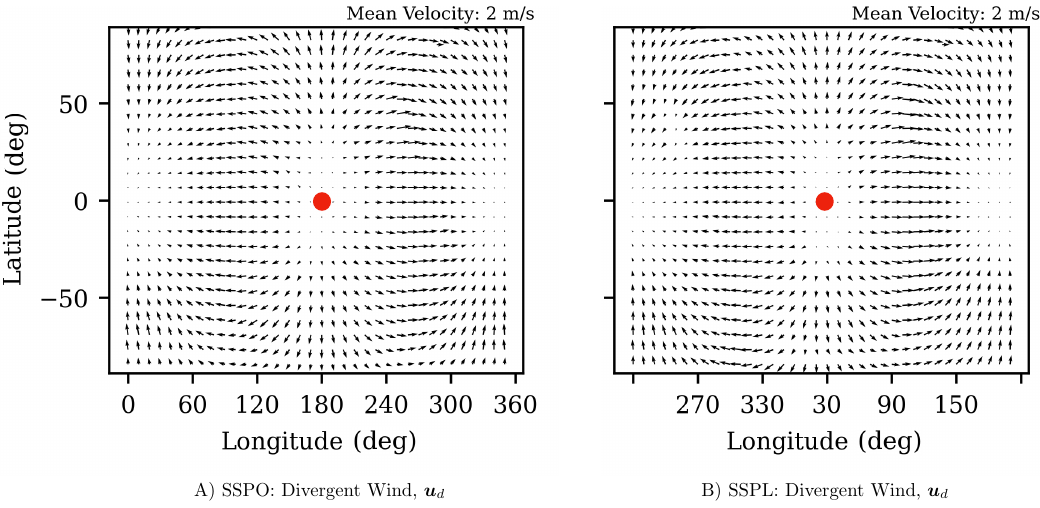}
\includegraphics[width=0.9\textwidth]{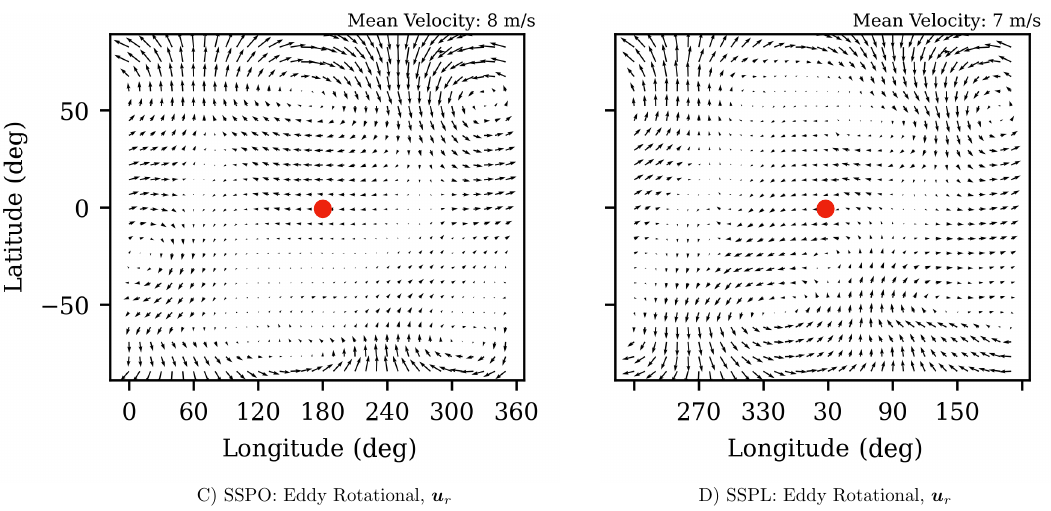}
\caption{Helmholtz decomposition of the vertically averaged horizontal winds for both our SSPO (left) and SSPL (right) models. The top row plots the divergent component of the wind ($U_{d}$) whilst the bottom row plots the eddy component of the rotational wind ($U_{e}$ - chosen to emphasises the standing wave structure which drives the zonal jets). The location of the sub-stellar point is marked with a red spot. \label{fig:mean_Helmholtz}}
\end{figure*}
In order to investigate the effects of the land-mass distribution and orography on the atmospheric dynamics, we turn to the Helmholtz wind decomposition \autoref{sec:helmholtz_method}, which isolates different components of the wind, each of which is has a different underlying driving mechanism. \\
We start by analysing the divergent (top row) and rotational (bottom row) components of the near surface horizontal wind (averaged over $P>10^{-1}$ bar), as shown in \autoref{fig:surface_Helmholtz_map}, in order to explore how the land-mass distribution and orography shape the dynamics. \\
Near the surface, the divergent component of the wind in both the SSPO and SSPL models (\autoref{fig:surface_Helmholtz_map}A/B) is dominated by a strong convergence at the sub-stellar point. This is the bottom of the day-side component of the global overturning circulation, which consists of heat rising on the day-side and sinking on the night side. Note that the top, divergent part of this upwelling can be seen in the radial mean divergent wind, as shown in \autoref{fig:mean_Helmholtz}. However, whilst this convergent wind dominates the profile, the effects of land/orography on the wind can also be felt. For example, consider the divergent winds over the mountainous regions of western North and South America, i.e. the American Cordillera. In North America we find that winds either break-up (SSPO) or form (SSPL) over the mountainous region, whilst in South America we find a consistent divergence from the narrow mountainous band in both models. Similar effects can be seen in the High-Mountain region of Asia centred over the Tibetan Plateau as well as in mountainous regions of Europe and Oceania, although the later is sensitive to the wind convergence at the sub-stellar point. 
We also find that Antarctica plays a significant role in shaping the winds in the southern hemisphere, driving significant divergent flows northwards. Again this can be linked to the orography of the region:
Antarctica is a highly mountainous and high altitude region (with an average elevation of $\sim2500$ m) which drives strong katabatic winds from the interior down the steep vertical drops along the coast. 
Briefly, katabatic winds are gravity driven flows of cold, dense, air parcels from { high altitudes to low. They occur because of radiative cooling of air parcels at high altitudes, which itself is driven by the relative coolness of high-altitude land-masses.}  On Earth, the strongest katabatic winds are associated with the ice-sheets of Antarctica and Greenland. \\
The effects of the land-masses/orography can also be seen in the rotational component of the near surface wind (\autoref{fig:surface_Helmholtz_map}C/D), which shows significant differences not only between the northern and southern hemisphere but also between flows over land and oceans. 
A good example of how the land-mass distribution shapes the rotational winds can be seen in Central America (SSPO) and South-East Asia (SSPL). These winds occur at a similar location with respect to the sub-stellar point in both the SSPO and SSPL cases, and in both cases we can see how the flow narrows and intensifies as it passes between two significant land-masses. Furthermore, if we compare the winds over either Asia or South America, we can see how the presence of orography/mountain ranges, both slows the wind and reshapes it, diverting it away from high altitude regions. 
Finally, we also find that significant circulations/vortices develop over the oceans (see the Indian, Pacific and Atlantic oceans) but these circulations are not symmetric and do not appear to form a standing wave as is typically found on tidally-locked exoplanets \citep{2011ApJ...738...71S}, suggesting that their driving is more localised than the global driving found with a slab ocean \citep{10.1093/mnras/stad2704}. \\

To investigate if this lack of a standing wave pattern is a robust feature of our model atmospheres or is instead linked to the influence of land/orography on the winds, we next explore the Helmholtz wind decomposition of the vertically averaged, over all pressure levels, horizontal wind.
Starting with the divergent wind (\autoref{fig:mean_Helmholtz}A/B), as previously alluded to, we see evidence for the global overturning circulation with a strong wind diverging from the sub-stellar point and converging on the night-side where it travels downwards. This wind profile is nearly identical in both the SSPO and SSPL models, and shows little-to-no asymmetry between the northern and southern hemispheres, suggesting that, away from the surface, land/orography has little effect on the divergent component of the wind, and instead it is the strong day-side irradiation which drives the divergent dynamics. \\ 
Moving onto the rotational component of the wind, we generally find that the profile is dominated by zonal-mean flows, such as the jets discussed in \autoref{sec:zonal_meridional}. As such, in order to visualise the driving forces behind these zonal jets as well as the effects of land/orography on the rotational winds, we further decompose the rotational wind into its zonal-mean and eddy components, plotting the latter component on the bottom row of \autoref{fig:mean_Helmholtz}. \\
Here we can see evidence of two distinct driving forces at play. 
The first is a standing-wave structure with two circulation cells in each hemisphere, although the exact strength of each cell as well as its extent is location and time dependent. For example, in the SSPO model we find two near-identical circulation cells in the northern hemisphere, whilst in the southern hemisphere we find a longitudinally broad anti-clockwise circulation cell on the day-side paired with a a smaller and weaker clockwise circulation cell on the night-side. A similar, but distinct, story holds true in the SSPL model. Note that the exact structure of the eddy wind is also highly dependent upon the pressure-level considered \footnote{This can be seen in the online version of these figures which shows the divergent and eddy wind components at each pressure level.}, leading to the asymmetric and pressure-dependent zonal-jet structure discussed in \autoref{sec:zonal_meridional}. \\ 
Together these changes suggest that the eddy component of the wind, and hence the rotationally driven dynamics (such as the jets discussed in \autoref{sec:zonal_meridional}) are much more sensitive to the effects of land and orography. Hence the deviation from the symmetric standing wave structure predicted by, for example, \citet{Williams1988,2011ApJ...738...71S} and found by \citet{2014MNRAS.445..930C,2018ApJ...852...67H} and \citet{10.1093/mnras/stad2704}. Examples of these orography induced changes include: the slower circulation cell south-east of the sub-stellar point in our SSPO model, a location which approximately corresponds to the location of Australia and the reduced circulation strength north of the sub-stellar point in our SSPL model, a region which approximately corresponds with Europe. { In general, these orography induced changes to the Rossby gyres and circulations are in good agreement with those found by \citet{2019AsBio..19...99D} for Proxima Centauri b.}

\subsection{Orography and Horizontal Winds} \label{sec:wave_breaking}
\begin{figure*}[tb]
\centering
\includegraphics[width=0.8\textwidth]{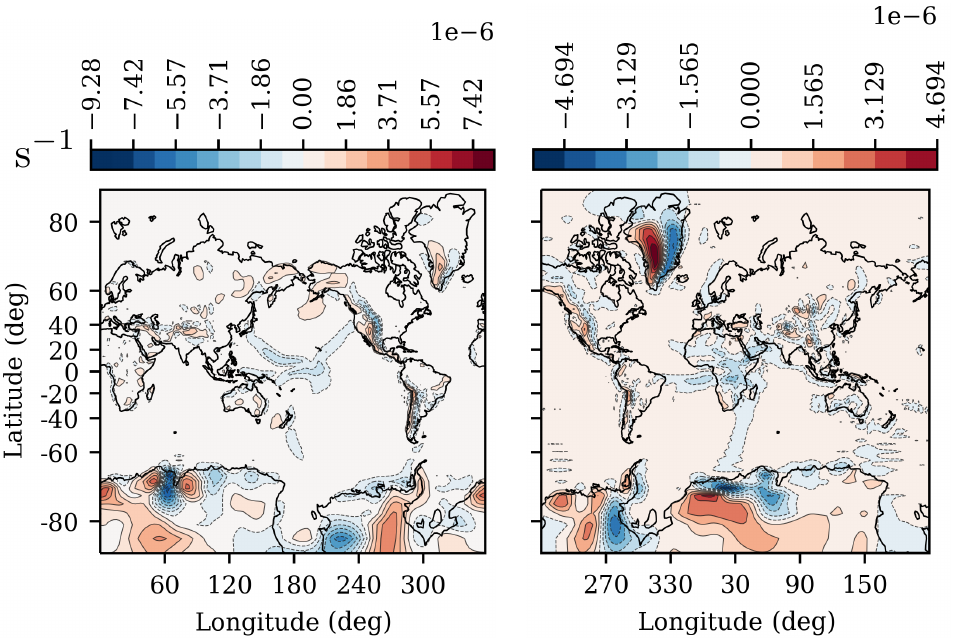}
\caption{Scalar divergence calculated from the vertically averaged near-surface horizontal winds for both our model with the sub-stellar point placed over the ocean (SSPO - left) and sub-stellar point placed over land/Africa (SSPL - right). The plot uses a Mercator projection to better resolve the divergence field over the Antarctic.  \label{fig:surface_divergence}}
\end{figure*}

To better visualise the link between orography and deviations from a symmetric circulation we finish our analysis of the dynamics by exploring the scaler divergence ($\nabla\cdot\bm{u}$) of the vertically averaged (over all $P>10^{-1}$ bar) near surface wind, as shown in \autoref{fig:surface_divergence}. \\ 

In general, the divergence of the surface wind traces a combination of the orography and the upwelling at the sub-stellar point, with the latter seen as a negative divergence at the centre of both panels of \autoref{fig:surface_divergence}. Both divergence profiles reveal the high-altitude land-masses discussed in \autoref{sec:helmholtz}, including winds diverging from the American Cordillera, chaotic mixing over the Tibetan Plateau, and mountain-ranges on Antarctica, and a previously unmentioned high-latitude mountainous region, Greenland, where we find a strong east-west divergence in the surface wind. Interestingly, the regions which show the largest divergence in the surface wind are also the regions that \citet{anand2024} and \citet{cooke2024} found that ozone accumulates in, suggesting an orographical explanation for their results. \\
The strong influence that orography has on the near surface winds, and in particular its ability to break the symmetry between winds in the northern and southern hemispheres of our otherwise near-symmetric models, suggest that future models of planetary atmospheres need to go beyond just considering the land-mass distribution. Instead, as we discuss in \autoref{sec:concluding_remarks}, models also need to consider both the land-mass distribution and the orography of the land. Further the possibility that land-masses are breaking the symmetry in atmospheric circulations must be considered when trying to understand unusual observations, such as the potential for asymmetric or concentrated ozone distributions \citep{10.1093/mnras/stad2704,cooke2024,anand2024}, or the enhancement of dis-equilibrium chemistry effects due to transport \citep{Chen_2018,Chen_2019}. It may also have a role in understanding potential habitability, with studies showing that orography might have shaped the global climate of the early Earth \citep{InfluenceofSurfaceTopographyontheCriticalCarbonDioxideLevelRequiredfortheFormationofaModernSnowballEarth,Walsh2019} or driven mass-extinction events \citep{Farnsworth2023}.  

\subsection{How Atmospheric Circulation Shapes the Surface Distribution of Ozone} \label{sec:surface_ozone}
\begin{figure}

\includegraphics[width=0.99\columnwidth]{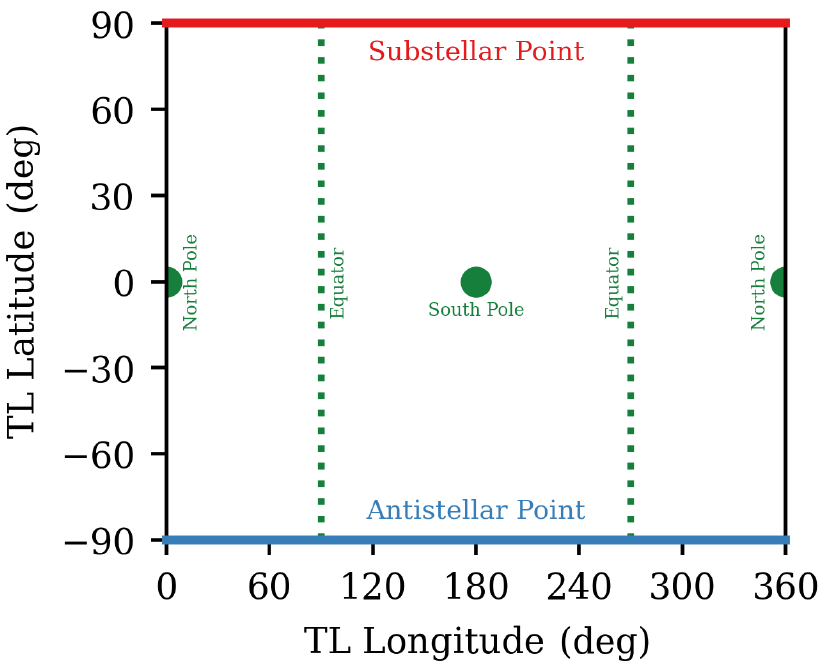}
\caption{ A schematic designed to aid the reader in interpreting horizontal maps plotted in tidally locked coordinates. Here tidally-locked latitude ($\theta_{TL}$) is a measure of the angle from the terminator, with the sub-stellar/anti-stellar point occurring at $\theta_{TL}=90^\circ$ and $\theta_{TL}=-90^\circ$ respectively, whilst each tidally-locked longitude ($\phi_{TL}$) represents an arc that connects the sub-stellar and anti-stellar points, passing through the north-pole at $\phi_{TL}=0^\circ$ and the south-pole at $\phi_{TL}=180^\circ$.   \label{fig:TL_coordinates_diagram}}
\end{figure}
\begin{figure*}[htp]
\begin{centering}
\includegraphics[width=0.75\textwidth]{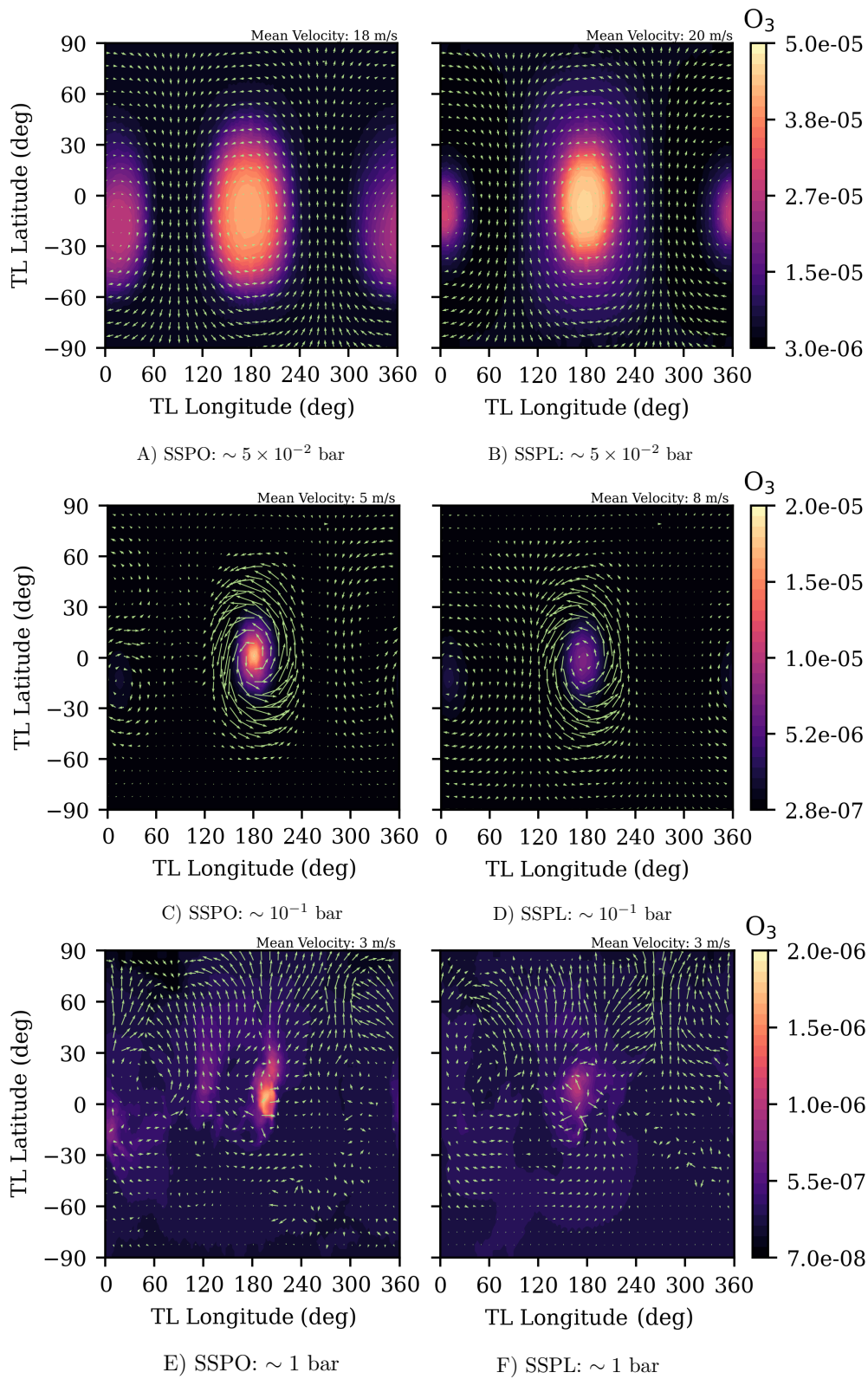}
\caption{ Horizontal slices of the fractional ozone (O$_3$) abundance at three pressure levels ($\sim5\times10^{-2}$ bar - top, $\sim10^{-1}$ bar - middle, and $\sim1$ bar - bottom), for both our model with the sub-stellar point placed over the ocean (SSPO - right) and the sub-stellar point placed over land/Africa (SSPL - right). Note that we have transformed the slices to tidally-locked coordinates in order to emphasise the build-up of ozone over the poles and the wind/vortices (shown using green quivers) which shape and constrain said build-up. \label{fig:TL_ozone}}
\end{centering}
\end{figure*}
We finish our analysis by investigating how the winds and circulations drive the aforementioned asymmetric ozone distribution found in our tidally-locked TRAPPIST-1e models. Specifically, as a coda to the work of \citet{anand2024}, we explore how cyclonic winds trap ozone at the south pole in both our SSPO and SSPL models. \\
To aid in this discussion, we transform our grid from the standard latitude-longitude grid used on Earth to the tidally locked coordinate system \citep{2015ApJ...802...21K,2021PNAS..11822705H} which emphasises flows between the sub-stellar and anti-stellar points and around the poles. 
Under this system, the tidally-locked latitude ($\theta_{TL}$) becomes a measure of the angle from the terminator, with the sub-stellar and anti-stellar points occurring at $\theta_{TL}=90^\circ$ and $\theta_{TL}=-90^\circ$ respectively. Whilst each tidally-locked longitude ($\phi_{TL}$) represents an arc connecting the sub-stellar and anti-stellar points, passing though the north-pole at $\phi_{TL}=0,360$ and the south pole at $\phi_{TL}=180$. This can be difficult to visualise at first, hence we provide a diagram in \autoref{fig:TL_coordinates_diagram} designed to aid the reader in interpreting the maps shown in \autoref{fig:TL_ozone}. We centre our figures on the south-pole. \\
As discussed in \citet{anand2024}, ozone forms aloft on the day-side before being transported to the night-side by a combination of the zonal jets and global overturning circulation (\autoref{fig:zonal_mean_zonal_wind} and \autoref{fig:mean_Helmholtz}). Here the ozone is transported not only down towards the surface but also towards the south-pole thanks to the asymmetric meridional circulation (\autoref{fig:meridional_circulation}), where it accumulates. 
But why exactly does it become trapped at the south pole and how can such high concentrations accumulate? The answer to the second question is rather simple: the obliquity of TRAPPIST-1e in both of our models is zero, which means that the poles are either dark or weakly illuminated (with any incoming irradiation having to pass through a thick column to reach the near-surface ozone - see \autoref{fig:stellar_insolation_maps}). As such the polar ozone is relatively stable, and a higher ozone concentration can be maintained (a similar effect occurs in the models of \citet{10.1093/mnras/stad2704} thanks to their ozone becoming trapped within night-side gyres). Yet this drop in insolation is not enough to explain the high concentrations found in our models (and in the work of \citet{cooke2024} and \citet{anand2024}) - much like the night-side gyres of \citet{10.1093/mnras/stad2704}, our results also require that some-kind of wind-structure confines the ozone to the poles, and in particular to the south pole. We suggest that this takes the form of a polar-vortex, with the strong orography of the Antarctic continent driving the enhanced circulation at the south pole. \\
Evidence for both this wind structure, as well as the role that orography plays in shaping it, can be seen in \autoref{fig:TL_ozone}. \\
Starting slightly further away from the surface (at $P\sim5\times10^{-2}$ bar - $\sim24$ km altitude - \autoref{fig:TL_ozone}A/B), we find vortices at both poles, centred on, and quiescent in, regions of peak ozone concentration. This suggests that the vortices are indeed acting as traps in which ozone accumulates. However differences between the vortices at the north and south poles, and between the SSPO and SSPL models are apparent. For example, we find that the vortices at the north-pole are generally weaker than those at the south-pole, reflecting the generally slower zonal wind speeds found in the northern hemisphere (\autoref{fig:zonal_mean_zonal_wind}). In turn we find that these weaker vortices are less effective at trapping ozone leading to a significant asymmetry between the ozone distribution at the north and south poles, with a much higher ozone concentration found at the south pole. We also find differences between the ozone distribution in our SSPO and SSPL models, although these differences are significantly smaller than that found between the northern and southern poles in both models. For example, whilst the peak ozone concentration at the south-pole is slightly lower in our SSPO than SSPL model, the total ozone concentration at this pressure level is $\sim15\%$ higher. This is because of differences in the distribution driven by differences in the relative configuration of the Antarctic land-mass in our models. In our SSPO model more of the Antarctic land-mass is on the night-side than the day-side, whereas the reverse is true in our SSPL model. In turn, the ozone at the south pole is generally closer to the anti-stellar point in our SSPO model, further insulating the ozone from any destructive ozone. 
Note that the differences between the ozone distribution in the SSPO and SSPL models is generally smaller at the north pole than the south pole, reflecting how the north pole is dominated by sea-ice in both models. \\
The asymmetry between the ozone distribution only grows as we move towards the surface ($P\sim1\times10^{-1}$ bar - $\sim13.5$ km altitude- \autoref{fig:TL_ozone}C,D) and the polar vortex at the north pole starts to weaken and break up. The vortex at the south pole also evolves, with the wind becoming further confined to high latitudes, shrinking the ozone peak and further concentrating ozone at the pole. Differences in the vortex at the south pole, and in the meridional transport more generally (see \autoref{sec:zonal_meridional}), then drive differences in the total ozone concentration at 0.1 bar. Not only is the ozone in our SSPO model more concentrated at the pole than in our SSPL model, with a 2x increase in peak ozone abundance, but the total ozone at this pressure level is $\sim22\%$ higher in our SSPO model. As for why these differences in polar circulations occur, the most likely cause is a combination of the Coriolis effect and Antarctic Katabatic winds: in general Coriolis forces will act to suppress off-equator flows, with the strength of this suppression growing as we move to higher latitudes, however at the south pole, where the polar plateau induces strong winds from high to low altitudes, we instead find that the Coriolis force acts to reshape the circulation, leading to the observed polar-vortex and north/south asymmetry. \\ 
Finally, near the surface ($P=1$ bar - $<1$ km altitude -  \autoref{fig:TL_ozone}E/F), we start to see evidence for the transport of ozone away from the south pole and back towards the equator. In our SSPO model this takes the form of a north-easterly wind transporting ozone from the south-pole on the night-side towards the equator on the day-side, whilst in our SSPL model the wind is more uniformly divergent, leading to the somewhat even transport towards the equator seen in \autoref{fig:TL_ozone}F.
Both patterns of ozone transport are compatible with the equatorward surface transport found in the zonal-mean meridional circulation (\autoref{fig:meridional_circulation}A/B), emphasising the need for a multi-dimensional analysis of the dynamics. Note that the differences in ozone abundance between out SSPO and SSPL models is still present near the surface, with the peak ozone abundance in our SSPO model being $\sim50\%$ higher than our SSPL model. However this difference is reduced to $\sim2\%$ in favour of our SSPO model when we integrate over the pressure level. \\
Overall we find that slight differences in the location of the Antarctic land-mass, and its associated orography, between our SSPO and SSPL models drive differences in the polar-vortex at the south-pole. When combined with the differences in atmospheric oxygen content driven the enhanced ocean evaporation in our SSPO model, this can help to explain the highly asymmetric and rich ozone distributions found in both our models and the models of \citet{cooke2024}.

\section{Concluding Remarks} \label{sec:concluding_remarks}

In this work we have used the Earth-System model WACCM6/CESM2 to simulate the atmospheric dynamics of the tidally-locked, and potentially habitable, terrestrial exoplanet TRAPPIST-1e assuming two different, Earth-like, land-ocean distributions: one in which the sub-stellar point is fixed over the Pacific ocean (SSPO) and one in which it is fixed over land, specifically central Africa (SSPL). The aim was to investigate how the presence of an Earth-like land-mass distribution, with its associated orography, affects the atmospheric dynamics and chemistry. For example, could the inclusion of Earth-like orography explain why both \citet{cooke2024} and \citet{anand2024} found an asymmetric accumulation of ozone over the south pole whereas \citealt{10.1093/mnras/stad2704} (who considered a model with a slab ocean), found a equatorially symmetric ozone distribution with ozone accumulating in off-equator vortices on the night-side. 
Here we consider models with two different sub-stellar point locations in order to investigate not only the effect of orography on the atmospheric dynamics, but also to distinguish between differences in the atmospheric dynamics associated with the presence of a land-mass at the sub-stellar point or due to differences in the land fraction between the northern and southern hemispheres (for Earth-like topography, 68$\%$ of the land-mass can be found in the northern hemisphere). { Note that a similar set of land-mass distributions for an Earth-like Proxima Centauri b was considered by \citet{2019AsBio..19...99D} who found dynamics that were broadly similar to our own, but did not consider a coupled chemistry model, and hence did not identify, for example, the effects of land-mass distribution on ozone.} \\
We started our analysis by exploring differences in the zonal-mean atmospheric composition of our two models. We found that {\it the composition of the two models is near indistinguishable, with the differences being primarily associated with the liquid-ocean fraction of the model}. These differences occur because, whilst TRAPPIST-1e is in the habitable zone of its host star, it's insolation is about two-thirds of the Earths, leading to much of its surface being frozen. As such, the only liquid-ocean is found near the sub-stellar point and since this region is dominated by a land-mass in one of our models (although even here we find liquid oceans near the coast on the day-side), we find significantly stronger ocean evaporation in our SSPO model than our SSPL model. This water acts as a source of atmospheric oxygen, increasing the relative abundance of oxygen carrying molecules (such as $O_3$, HO$_2$, NO$_2$ etc.), particularly at low pressures where water photodissociates.
Note that, the above only holds true because of the Earth-like atmospheric composition of our models. If we were to consider different atmospheric compositions, particularly compositions with increased greenhouse gases, it is possible that both models would be warm enough to maintain significant liquid oceans away from the sub-stellar point \citep{2017ApJ...839L...1W}. However the tidally locked nature of the insolation means that differences in ocean-surface evaporation rates are likely to persist due to differences in ocean insolation. \\
We also find that the zonal-wind and meridional circulation profiles are broadly similar between our SSPO and SSPL models. 
{\it Analysis of the zonal-mean zonal-wind reveals that differences between the winds/jets in the northern and southern hemisphere are generally larger than the differences between our SSPO and SSPL models, with the asymmetry peaking in the off-equator jets found near the surface ($\sim0.5$ bar)}. A similar story also holds true for the zonal-mean meridional circulation. {\it In both our SSPO and SSPL models we find a meridional circulation profile which is dominated by a single cell in each hemisphere, a circulation cell which is reminiscent of the Hadley cells but which extends from the equator to the pole. These circulation cells combine to drive a net upflow slightly north of the equator}, similar to the seasonal shift in the Hadley-cell convergence zone on Earth. { The presence of a single, rotationally influenced, meridional circulation cell (Hadley-cell) per hemisphere is similar to the results reported by, for example, \citet{Williams1988,1988ClDy....3...45W}, \citet{Navarra2002}, \citet{2014MNRAS.445..930C,2015MNRAS.453.2412C}, \citet{2018ApJ...852...67H} and \citet{2018GeoRL..4513213G} amongst others. }
Whilst the circulation structure becomes more complicated when we confine our zonal averages to near the sub-stellar and anti-stellar points, we still find that the differences between the SSPO and SSPL circulations are small. Furthermore, near the surface, these differences appear to be highly correlated with the land-mass distribution. For example, at the sub-stellar point, we find two circulation cells in each hemisphere, with the location of the switch from the Hadley-like cell near the equator and the Ferrel-like cell near the pole occurring at approximately the same latitude that both the SSPO and SSPL models go from being locally land-dominated to ocean-dominated. Note however that, as we discuss above, the ocean is frozen away from the sub-stellar point. And since it is likely that the interactions between the near-surface wind and sea-ice will be differ from the interaction with a dynamic liquid ocean, it is possible that a different circulation pattern will be found for a hotter planet. This is something we intend to investigate as part of a future study. \\
The above differences in the winds and circulations between the northern and southern hemispheres suggested that orography might be playing a significant role in shaping the near surface winds. We investigated this premise in more detail by exploring the Helmholtz-wind decomposition of both the near-surface and vertically averaged winds. 
{\it Near the surface, we found that both the divergent and rotational components of the wind revealed significant shaping by land-masses, particularly the presence of orographic features, such as the American Cordillera, the Tibetan Plateau, or the Antarctic mountains}. { Note that similar continental wind shaping was found by \citet{2019AsBio..19...99D} in their Earth-like Proxima Centauri b models.} These high-altitude regions also act as sources of wind, for example the katabatic winds which flow from high altitudes to low, winds which can influence the atmospheric composition. {\it The effects of orography of the winds is also present far from the surface, with even the lowest pressure regions of our atmospheres revealing at least a hint of north-south asymmetry which is correlated with the land-mass distribution and its associated orography}. For example, the eddy component of the vertically averaged rotational wind revealed significant differences in the zonal-jet driving standing Rossby and Kelvin wave pattern between both our SSPO and SSPL models, and between the northern and southern hemispheres in both models. \\
{\it The effects of orography on the near-surface winds and circulations may also help to explain why ozone primarily accumulates at the south pole in both our models and the models of \citet{cooke2024} and \citet{anand2024}}: as we approach the surface and the radiative forcing weakens, the Coriolis effect is generally able to suppress high-latitude flows. However at the south-pole there is another source of winds, the Antarctic katabatic winds associated with the sharp vertical descent between the Antarctic continent and the surrounding ocean. Rotation reshapes these winds into a vortex which confines ozone to the south pole (or more specifically over the land-mass) almost all the way down to the surface where it then travels equatorward, leading to the lethal surface abundances discussed in \citet{cooke2024}. Further, differences in the orientation of the Antarctic land-mass with respect to the sub-stellar point in our SSPL and SSPO model, combined with the relative oxygen content driven by ocean-evaporation in our SSPO model, lead to differences in the ozone concentration between $\sim10^{-2}$ bar and the surface. For example, {\it we found that at $\sim10^{-1}$ bar the peak ozone concentration in our SSPO model was approximately twice that found in our SSPL model, and even the integrated ozone abundance was $\sim22\%$ higher than that found in our SSPL model.}  \\

Whilst the above results are interesting, it is important to remember that WACCM6/CESM2 as a model is highly tuned to the Earth's atmosphere. Therefore, care should be taken as the results presented here, and in other work with WACCM6/CESM2, may not be generally applicable. 
However that does not mean that such models cannot inform us about how heretofore unconsidered planetary features (such as the presence of a dynamic ocean with a complex land-mass distribution instead of a highly simplified water-world with a shallow, slab, ocean and no symmetry breaking orography) might affect planetary atmospheric dynamics, chemistry, and potentially habitability, and hence observations. \\
As such, we suggest that additional development time should be assigned to developing flexible land models which can be coupled with complex GCMs like WACCM6/CESM2. These flexible land models should allow us to not only explore how changing the land-mass fraction between the day-side and night-side and between the northern and southern hemispheres can break the global symmetry in circulations but also to investigate the effects of orography and land-surface composition on the dynamics in more detail. For example, how might the presence of Mars-like mountains (such as Olympus Mons) on a low-mass terrestrial exoplanet affect the dynamics and chemistry, and would such changes be large enough to be observable with future missions? Does the generally low elevation and ocean free topography of Venus shape the atmospheric dynamics in a way that is distinct from the Earth?
Could cratering and the formation of massive canyons also have a noticeable effect on the dynamics? Or could the presence of a super-continent, such as that found in both Earth's past and potentially in Earth's future \citep{2018GPC...169..133D}, drive terrestrial atmospheres away from or towards habitability \citep{2021GGG....2209983W}? In essence, both other planets in our solar system and our own Earth reveal diverse surfaces and dynamics, and there is no reason to not to expect that this will hold true for extra-solar planets.  \\
{\it Therefore it is safe to say that, as shown by our models, understanding how such land-masses and orography influence the global atmospheric dynamics and hence chemistry may be key to interpreting future observations. This includes assessing if a planet is truly habitable, or if, for example, instead some quirk of the dynamics means that significant, observable ozone (a potential biosignature) can accumulate even when the oxygen content is significantly reduced \citep{2023ApJ...959...45C}.  }

\begin{acknowledgements}
\nolinenumbers
F. Sainsbury-Martinez and C. Walsh would like to thank UK Research and Innovation for support under grant number MR/T040726/1. Additionally, C. Walsh would like to thank the University of Leeds and the Science and Technology Facilities Council for financial support (ST/X001016/1). This work was undertaken on ARC4, part of the High Performance Computing facilities at the University of Leeds, UK.\\
\end{acknowledgements}

\bibliography{papers}{}
\bibliographystyle{aasjournal}

\end{document}